# Classic papers: *déjà vu*, a step further in the bibliometric exploitation of Google Scholar

**Emilio Delgado López-Cózar, Alberto Martín-Martín, Enrique Orduna-Malea**

*EC3 Research Group: Evaluación de la Ciencia y de la Comunicación Científica, Universidad de Granada (Spain)*

## ABSTRACT

After giving a brief overview of Eugene Garfield's contributions to the issue of identifying and studying the most cited scientific articles, manifested in the creation of his *Citation Classics*, the main characteristics and features of Google Scholar's new service -Classic Papers-, as well as its main strengths and weaknesses, are addressed. This product currently displays the most cited English-language original research articles by fields and published in 2006.

## KEYWORDS

**Highly cited papers / Most cited papers / Citation Classics / Classic papers / Citation counts / Citation analysis / Bibliometrics / Scientometrics / Google Scholar**

14 pages, 6 tables, 6 figures

# 1. The Precursor: Eugene Garfield's Citation Classics

On the 3rd of January of 1977, exactly forty years ago, **Eugene Garfield** started to publish what he then called *Citation Classics*, a collection of short essays that featured the top 500 most cited articles published between 1961 and 1975 (Figure 1). From that moment until 1993, the *Current Contents* service published no less than 400 "*Citation Classic Commentaries*"[1]. The intention of these pieces was to present "the human side of scientific reports" through comments from the very researchers that had published them: how they came to be, who collaborated in the process, the problems that occurred during their development, the obstacles that were faced, and how the results were received by their colleagues.

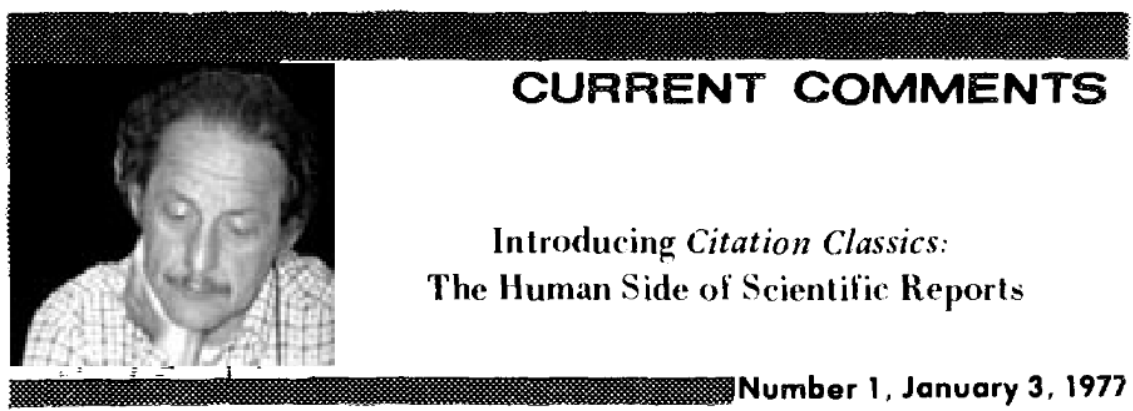

*Figure 1. Creation of the Citation Classics. Eugene Garfield. Introducing Citation Classics: The human side of scientific papers. Current Contents, 3 January 1977, 1, p. 5-7.*

But this idea was not new, because in 1969 **Garfield** had already compiled a list of the top 50 most cited articles published in 1967 (Figure 2). In that list he already used the term "*classics*" to refer to those highly cited documents. Six years later he prepared a similar list, but this time about articles published between 1961 and 1972 (Figure 3). This list comprised the top 50 most cited articles published in that period, and he again used the term "*classics*" to refer to those works.

---

[1] All of them available from http://garfield.library.upenn.edu/classics.html

*Fig. 1.* Fifty most cited articles for 1967, ranked according to total times cited. (Refer to Appendix A)

**Figure 2. Most cited articles published in 1967. Eugene Garfield. Citation indexing, historio-bibliography and the sociology of science. Current Contents, 14 April 1971, 6.**

**Garfield** revisited this topic repeatedly in the following years. No less than 17 essays about the "*citation classics*"[2] of various scientific fields or journals were published, and some of them stimulated a discussion on the meaning and influence of this kind of studies (immortality, obliteration, productivity, genre, Nobel prizes). Other essays (more than 80) were dedicated to examining the most cited papers, books, and authors in various disciplines, specialties, journals, or countries.

In short, a mammoth task that speaks volumes about **Garfield's** personality, a person who was ahead of his time in this topic and many other topics related to information retrieval and scientific evaluation. **Garfield**, recently deceased, was a pioneer whose memory we should always honour. These words are a tribute to him, but they are also a way to contextualize the birth of a new product. Nothing happens in a vacuum. We are always riding on the shoulders of giants… **Garfield** was the forefather of everything we do today, and of course, the precursor that enabled the creation of services like *Google Scholar*, and therefore partly responsible for the way we nowadays discover, retrieve, and evaluate scientific information.

[2] All of them available at http://garfield.library.upenn.edu/citationclassicsessays.html

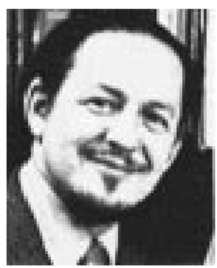

"current comments"

Selecting the All-Time Citation Classics. Here Are the
Fifty Most Cited Papers for 1961-1972.

**January 9, 1974**                                                    **Number 2**

About six years ago we compiled a list of the 50 papers most cited in the *Science Citation Index* ® in 1967.[1] Re-yet peaked; its 1972 count of 357 was up 20% from 1971.

Despite their 'age' many of the

*Figure 3. Top 50 most cited articles published between 1961 and 1972. Eugene Garfield. Selecting the All-Time Citation Classics. Here Are the Fifty Most Cited Papers for 1961-1972. Current Contents, 9 January 1974, 2, p. 5-8*

On top of this foundations, *Thomson Scientific* first, *Thomson Reuters* later, and today *Clarivate Analytics*, built the *Essential Science Indicators*, which every year presents the most cited documents of the last decade[3].

## 2. What does Google Scholar's Classic Papers offer?

The top 10 most cited English-language original research articles published in 2006 in each of 252 subject categories, according to the data available in Google Scholar as of May 2017. The total number of articles displayed in the product is 2515 articles[4].

### 2.1. Inclusion/Exclusion criteria

In order to make it to this product, articles must meet the following criteria:

- They must have been published in 2006
- They must be journal articles, articles deposited in repositories, or conference communications.
- The documents must describe original research. Review articles, introductory articles, editorials, guides, commentaries, etc. are explicitly excluded.
- They must be written in English.
- They must be among the top 10 most cited documents in their respective subject category.
- They must have received at least 20 citations.

---

[3] https://images.webofknowledge.com/images/help/WOS/hs_citation_applications.html
[4] Google Scholar's Classic Papers published in 2006. https://doi.org/10.13140/RG.2.2.27340.62084

## *2.2. Layout and visualization*

Articles are classified in 294 subject categories, which in turn are grouped in eight broad scientific areas (Table 1). However, there are 42 subject categories that appear in two broad scientific areas. Thus, there are 252 unique subject categories.

| Areas | Number of subject categories |
|---|---|
| Health & Medical Sciences | 68 |
| Engineering & Computer Science | 57 |
| Social Sciences | 51 |
| Life Sciences & Earth Sciences | 38 |
| Humanities, Literature & Arts | 25 |
| Physics & Mathematics | 23 |
| Chemical & Material Sciences | 17 |
| Business, Economics & Management | 15 |

*Table 1. Number of subject categories in each broad scientific area in Google Scholar's Classic papers.*

Each of these 252 categories presents 10 articles, except *French Studies*, which only has 5 (because they could not find more than 5 articles with at least 20 citations, which is the self-imposed minimum used by *Google Scholar*). That is the reason why the total number of articles is 2515 instead of 2520 (252 times 10).

For each article, the information displayed is:

- *Title of the study*, with a hyperlink to the record of the document in *Google Scholar*
- *Name of the authors*. Not all of them are displayed, only the ones that can fit in about 50 characters. For those authors that have set up a public *Google Scholar Citations* profile, the name is underlined and a link is available to said profile
- *Name of the journal, conference, or repository*, where the article has been published
- *Number of citations*
- *Picture and hyperlink to the Google Scholar Citations profile of one of the authors*, if available. If there are several co-authors with a profile, the system gives preference to the first author, then to the last author, and if neither of these have a profile, it selects whatever profile is available first, by author order.

This product, as could not otherwise, has the identifying traits of most of Google's products:

- **Simple and straightforward**: a list of the most cited articles in each discipline, with a simple browsing interface.
- **Easy to use and understand**: organized by broad scientific areas and inside of them by subject categories. Three clicks are enough to reach the documents or the public *Google Scholar Citations* profiles of their authors.

- **Minimal information**: As a whole, the product displays just over 2500 highly cited articles. Each article presents the most basic bibliographic information.
- **Little methodological transparency**: It is common for *Google Scholar* not to declare in detail how their products are developed.

Regarding the las point, there are four critical aspects about which we should know more precise information. They are aspects that could compromise the reliability and validity of the product:

**The first** of them is related to what *Google* understands as a research article. Although they declare that they are "…articles that presented new research"[5], we ask: how have they identified research articles from those that are not research articles? What constitutes an introductory article and how have they identified them? What do they mean when they add a disconcerting "etc." when they list the excluded document types? "Etc." is rarely admissible in science, where all explanations should be precise. This issue is important because it may be the case that some articles that don't meet these requisites have been included, or the opposite, that some articles that do meet the requisites are missing.

Actually, some *Twitter* users have already denounced that there are highly cited documents missing from their respective categories (Figure 4). Although the third article mentioned by *Twitter* user @TrevorABranch was published in an issue called "Reviews in Fish Immunology"[6] and the second one is classified as a report by *Science*[7], which might explain why they have been excluded (neither has being considered an original research article), the first one is indeed classified as a research article by *Science*[8], and has more citations than any article in that category (3,223 citations), which makes us wonder about the specific criteria used by *Google Scholar* to define the typology of the documents.

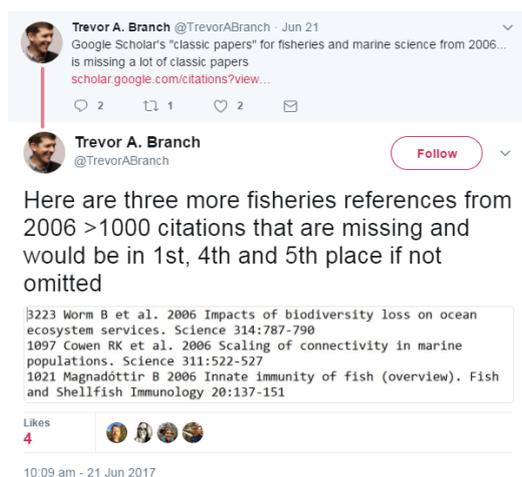

***Figure 4. Papers missing from the category "Marine Sciences & Fisheries" according to Twitter user @TrevorABranch***

---

[5] https://scholar.googleblog.com/2017/06/classic-papers-articles-that-have-stood.html
[6] http://www.sciencedirect.com/science/article/pii/S1050464805000781
[7] http://science.sciencemag.org/content/311/5760/522
[8] http://science.sciencemag.org/content/314/5800/787

It is important to remember that defining the typology of a document is not an easy task, and that even traditional bibliographic databases like *Web of Science* or *Scopus* have not been able to solve this issue completely. There are many discrepancies in how each of these databases defines the typology of the documents they cover. This happens frequently with review articles. There are also abundant internal inconsistencies in the databases.

**The second aspect** has to do with the subject classification of the articles. This task involves assigning each article to one of 252 subject categories, and it is a crucial issue for the correct development of the product, but also very thorny. There are two fundamental questions we may ask regarding this issue:

a) **Which criteria have they adopted to carry out the subject classification?**

This is a question we already asked in our previous analyses of *Google Scholar* Metrics[9]. It seems clear that the classification scheme they have selected is the same they use in *Google Scholar Metrics*, their annual ranking of scientific journals. The only difference is the elimination of eight subject categories. All of them share something in common: they are the categories referred to as "general", because their title is the same as the broad scientific area where they are included:

- Physics & Mathematics (general)
- Business, Economics & Management (general)
- Chemical & Material Sciences (general)
- Health & Medical Sciences (general)
- Engineering & Computer Science (general)
- Life Sciences & Earth Sciences (general)
- Social Sciences (general)
- Humanities, Literature & Arts (general)

At first, the elimination of these categories should not pose any problem, because the journals included in those categories are also classified in other subject categories (sometimes up to four other). However, there are journals which are only classified in these generic categories. Have the articles published in these journals been classified in other subject categories?

We have checked that articles published in multidisciplinary journals (such as *Nature*, *Science*, or *PNAS*) have been indeed classified *ad hoc* in their respective subject categories according to the topic of the articles. It seems that the articles published in journals with a broad scope have also been classified in the correct subject categories (Journal of the American Chemical Society, *IEEE Transactions on Industrial Electronics, The New England Journal of Medicine, JAMA, The Lancet, Qualitative Inquiry,*

[9] Martín-Martín, A., Ayllón, J. M., Orduña-Malea, E., & Delgado López-Cózar, E. (2014). Google Scholar Metrics 2014: a low cost bibliometric tool. EC3 Working Papers, 17. arXiv preprint arXiv:1407.2827.

*Scientific Reports, PLoS Biology, Reviews of Modern Physics, Procedia-Social and Behavioral Sciences*). This issue makes us wonder a second question…

### b) How have they classified the articles published in multidisciplinary journals and journals with a broad scope?

Considering that most services rely on journal-level classifications and not on article-level classifications, how has *Google Scholar* solved this problem? In most cases articles are simply assigned to the same categories where the journal has been classified, without paying attention to the actual topic of the article.

This approach, the most commonly used in bibliometrics, is ill-suited for multidisciplinary journals and the other journals with a broad scope that are published in most disciplines. We know that the *Essential Science Indicators* (ESI) classifies multidisciplinary articles according to the subject categories of the journals publishing the articles that cite them as well as to the journals of the articles cited by them, an incontrovertible approach. Therefore, how has *Google Scholar* done this?

**The third aspect** has to do with another crucial issue related to the way *Google Scholar* works: can we be sure they have successfully merged together all the versions indexed in *Google Scholar* of these documents? Otherwise, the citation counts of the documents might be scattered in several records.

In previous studies we have shown that this is an important issue when we are talking about highly cited articles[10]. It seems, as Figure 5 shows, that there are still some records that refer to the same highly cited documents that appear in *Classic Papers* which haven't been merged with the main record (the one with the most citations).

---

[10] Martín-Martín, A., Orduña-Malea, E., Ayllón, J. M., Delgado López-Cózar, E. (2014). Does Google Scholar contain all highly cited documents (1950-2013)? EC3 Working Papers, 19. arXiv preprint arXiv:1410.8464

**Figure 5. Examples of documents for which there are several versions that have not been properly merged to the main record.**

**The fourth aspect** has to do with the threshold selected to consider an article a "*classic paper*". Why did they decide to set this number to 10 articles in each subject category? Why is this threshold the same for the 252 subject categories?

This decision goes against logic and long-established bibliometric practices, where the different natures of the various scientific disciplines have long been acknowledged. Different scientific communities have different citation habits and different sizes in terms of number of researchers. In order to illustrate this inconsistency, Table 2 shows the 10 WoS categories with the highest number of papers published in 2006, and the 10 categories with the lowest number of papers published in the same year. Next to the number of papers, another column shows the fraction that 10 articles is respect to the total amount of articles in the category.

| Web of Science Categories | N papers | % covered by 10 documents |
|---|---|---|
| Engineering Electrical Electronic | 86,568 | 0.012 |
| Computer Science Artificial Intelligence | 61,137 | 0.016 |
| Materials Science Multidisciplinary | 53,671 | 0.019 |
| Physics Applied | 49,267 | 0.020 |
| Biochemistry Molecular Biology | 47,259 | 0.021 |
| Chemistry Physical | 39,715 | 0.025 |
| Telecommunications | 37,641 | 0.027 |
| Computer Science Theory Methods | 36,233 | 0.028 |
| Optics | 33,660 | 0.030 |
| Physics Condensed Matter | 32,806 | 0.030 |

| Web of Science Categories | N papers | % covered by 10 documents |
|---|---|---|
| Psychology Mathematical | 498 | 2.008 |
| Primary Health Care | 484 | 2.066 |
| Medical Ethics | 474 | 2.110 |
| Dance | 401 | 2.494 |
| Literature American | 399 | 2.506 |
| Andrology | 378 | 2.646 |
| Poetry | 368 | 2.717 |
| Literature Slavic | 254 | 3.937 |
| Folklore | 205 | 4.878 |
| Literature African Australian Canadian | 175 | 5.714 |

*Table 2. Number of papers classified in the 10 most productive (top) and least productive (down) WoS categories*

While in *Engineering Electrical Electronic* and *Computer Science Artificial Intelligence* those 10 documents make up barely 0.01% of the total, in *Folklore* and *Literature African Australian Canadian*, 10 articles make up more than 5% of the articles in the category.

This productive disparity among disciplines goes together with also huge differences in citation patterns. Table 3 displays the maximum and minimum number of citations in the 10 articles displayed in *Classic Papers* in the 15 categories with highest (top) and lowest (down) number of citations. This way it is easy to see the problem of selecting the same citation threshold (20) for all subject categories.

| Subcategories | Citations (10 most cited articles) | | |
|---|---|---|---|
| | Maximun | Minimun | Total |
| Information Theory | 18,648 | 1,179 | 51,987 |
| Psychology | 29,294 | 1,181 | 42,226 |
| Cell Biology | 17,121 | 1,278 | 36,359 |
| Oncology | 6,987 | 2,411 | 35,763 |
| Bioinformatics & Computational Biology | 9,981 | 1,555 | 34,680 |
| Condensed Matter Physics & Semiconductors | 8,415 | 1,640 | 34,379 |
| Immunology | 5,706 | 1,706 | 23,200 |
| Economics | 3,112 | 1,883 | 23,048 |
| Molecular Modeling | 9,745 | 766 | 22,823 |
| Astronomy & Astrophysics | 6,624 | 1,056 | 21,854 |
| Finance | 2,958 | 1,065 | 21,496 |
| Psychiatry | 3,059 | 1,313 | 20,127 |
| Atmospheric Sciences | 2,763 | 1,319 | 19,684 |
| Biophysics | 4,556 | 760 | 19,610 |
| Cardiology | 2,824 | 1,378 | 18,853 |

| Subcategories | Citations (10 most cited articles) | | |
|---|---|---|---|
| | Maximun | Minimun | Total |
| Religion | 300 | 102 | 1,743 |
| History | 341 | 104 | 1,682 |
| Economic History | 328 | 81 | 1,586 |
| Latin American Studies | 231 | 103 | 1,396 |
| Bioethics | 237 | 90 | 1,272 |
| Literature & Writing | 353 | 72 | 1,263 |
| Visual Arts | 155 | 89 | 1,101 |
| Film | 536 | 37 | 1,049 |
| Technology Law | 75 | 41 | 1,014 |
| European Law | 178 | 63 | 978 |
| Middle Eastern & Islamic Studies | 225 | 58 | 966 |
| Canadian Studies & History | 182 | 42 | 706 |
| American Literature & Studies | 81 | 32 | 545 |
| Drama & Theater Arts | 69 | 34 | 450 |
| French Studies | 32 | 20 | 131 |

*Table 3. Citations in the 15 subject categories in Classic Papers with highest (top) and lowest (down) numbers of citations overall.*

There is no one better than **Eugene Garfield** to highlight this reality since he acknowledges this problem when discussing what a "*citation classic*" is. He said "Citation rates differ for each discipline. The number of citations indicating a classic in botany, a small field, might be lower than the number required to make a classic in a large field like biochemistry. In general, a publication cited more than 400 times should be considered a classic; but in some fields with fewer researchers, 100 citations might qualify a work"[11].

The *Highly Cited Papers* available in the *Essential Science Indicators* (currently owned by *Clarivate Analytics*), follow the same principles delineated by **Garfield**. Today the product "lists the top cited papers over the last 10 years in 22 scientific fields. Rankings are based on meeting a threshold of the top 1% by field and year based on total citations received. Citation cutoffs specific to field and year are applied to all papers in the journal set to select highly cited papers. Citation thresholds are based on the distribution of citations, picking the specified top fraction of papers for each year and field. The thresholds are based on the cutoffs given in the All Years column of the Baseline Percentiles table"[12].

One of the most innovative aspects of the product is that it displays the link to the *Google Scholar Citations* profile of some of the authors of the article. 654 of the 2515 articles (31%) displayed in "*Classic papers*" lack such a link, and there are significant differences among disciplines. For example, in Chemical & Material Sciences, 5 out of the 17 subdisciplines considered (0.29%) display links to author profiles for all documents included in the subdiscipline, whereas in Humanities, Literature & Arts, in none of the 25 subcategories can we find at least one author with a public profile for each of the 10 documents (Table 4).

| Category | Subcategories | SWP | % |
|----------|--------------:|----:|-----:|
| Life Sciences & Earth Sciences | 38 | 7 | 0,18 |
| Business, Economics & Management | 15 | 4 | 0,27 |
| Chemical & Material Sciences | 17 | 5 | 0,29 |
| Engineering & Computer Science | 57 | 15 | 0,26 |
| Humanities, Literature & Arts | 25 | 0 | 0,00 |
| Health & Medical Sciences | 68 | 6 | 0,09 |
| Physics & Mathematics | 23 | 3 | 0,13 |
| Social Sciences | 51 | 5 | 0,10 |
| TOTAL | 294 | 45 | |

***Table 4. Number of subcategories in which all documents are linked to at least one Google Scholar Citations profile***
Note: NWP: Number of subcategories with at least one author profile linked

---

[11] Garfield, E. Short History of Citation Classics Commentaries. Available at http://garfield.library.upenn.edu/classics.html
[12] https://images.webofknowledge.com/images/help/WOS/hs_citation_applications.html

The subcategories in which all 10 highly cited documents have at least an author with a *Google Scholar Citations* profile are listed in Table 5.

| |
|---|
| Artificial Intelligence, Computer Graphics, Computer Networks & Wireless Communication, Computer Vision & Pattern Recognition, Data Mining & Analysis, Databases & Information Systems, Multimedia, Software Systems, Human Computer Interaction |
| Economics, Entrepreneurship & Innovation, Business, Economics & Management, Human Resources & Organizations, Game Theory and Decision Science, Probability & Statistics with Applications |
| Biodiversity & Conservation Biology, Sustainable Development, Urban Studies & Planning, Environmental Sciences, Atmospheric Sciences, Genetics & Genomics, Developmental Biology & Embryology, Evolutionary Biology, Biochemistry, Ocean & Marine Engineering |
| Inorganic Chemistry, Polymers & Plastics, Materials Engineering, Electromagnetism, Nanotechnology, Structural Engineering, Quantum Mechanics |
| Developmental Disabilities, Pulmonology, Psychiatry, Rehabilitation Therapy |
| Political Science, Family Studies |

*Table 5. Subcategories in Classic Papers in which there is at least one author of the articles displayed who has a public Google Scholar Citations profile.*

Lastly, Table 6 shows the subcategories in which there is a higher number of highly cited documents for which no author profile is available. As we can observe, American Literature & Studies and, unexpectedly, Plastic & Reconstructive Surgery, are at the top of this list.

| Subcategories | Number of papers for which no author has a public GSC profile |
|---|---|
| American Literature & Studies | 9 |
| Plastic & Reconstructive Surgery | 9 |
| Drama & Theater Arts | 8 |
| International Law | 8 |
| African Studies & History | 7 |
| Dentistry | 7 |
| Ethnic & Cultural Studies | 7 |
| Literature & Writing | 7 |
| Visual Arts | 7 |

*Table 6. Subcategories in Classic Papers in which most of the documents are written by authors that haven't set up a public Google Scholar Citations profile.*

Most of the articles displayed in "*Classic Papers*" are written in collaboration by several co-authors, and even if more than one has a public *Google Scholar Citations profile*, only one is prominently displayed in the record. The system seems to give preference to the first author, then to the last author, and if neither of these have a profile, it selects whatever profile is available first according to author order.

## 2.3. The surprise

Surprisingly for a Google product: there is no search feature. The search box is absent, and therefore users cannot search articles using keywords. For the first time, Google Scholar forces us to use browsing as the only way to navigate the information available in the product. Users will have to first select a broad category, then a subcategory, and then they'll be presented with the 10 most cited articles of that subcategory. Additionally, they have also changed the interface of Google Scholar Metrics, which now has an even more minimalist feel, and they have grouped Classic Papers and the journal lists under the same tag "METRICS" (Figure 6). Is this a sign of more future changes in Google Scholar's products? We cannot know for sure, and we'll have to wait until summer, the season when Google Scholar usually releases its innovations.

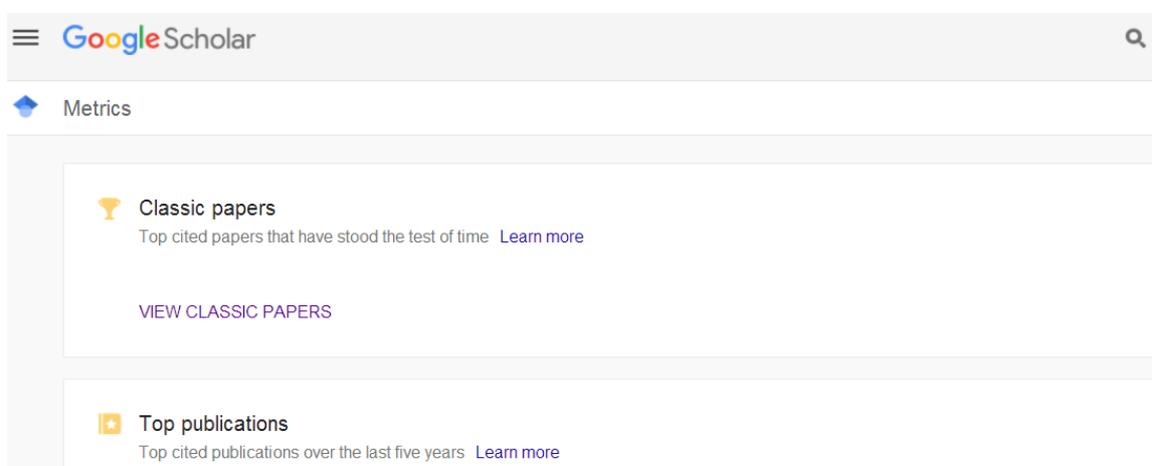

*Figure 6. New interface of Google Scholar Metrics: Access to Citation Papers and Top publications*

## Acknowledgments

Alberto Martín-Martín enjoys a four-year doctoral fellowship (FPU2013/05863) granted by the Ministerio de Educación, Cultura, y Deportes (Spain). Enrique Orduna-Malea holds a Juan de la Cierva postdoctoral fellowship (IJCI-2015-26702) by the Ministerio de Economía y Competitividad (Spain).

# REFERENCES

Garfield, E. (1971). Citation indexing, historio-bibliography and the sociology of science. *Current Contents*, 6, pp. 156-157.

Garfield, E. (1974). Selecting the All-Time Citation Classics. Here Are the Fifty Most Cited Papers for 1961-1972. *Current Contents*, 2, pp. 5-8.

Garfield, E. (1977). Introducing Citation Classics: The human side of scientific papers. *Current Contents*, 1, pp. 5-7.

Martín-Martín, A.; Ayllón, J. M.; Orduna-Malea, E. & Delgado López-Cózar, E. (2014). Google Scholar Metrics 2014: a low cost bibliometric tool. *EC3 Working Papers*, 17. Online available at: https://arxiv.org/abs/1407.2827

Martín-Martín, A.; Orduna-Malea, E.; Ayllón, J. M. &  Delgado López-Cózar, E. (2014). Does Google Scholar contain all highly cited documents (1950-2013)?  *EC3 Working Papers,* 19. Online available at: https://arxiv.org/abs/1410.8464